# Enhancing Infant Crying Detection with Gradient Boosting for Improved Emotional and Mental Health Diagnostics

*Kyunghun Lee*[⋆] *Lauren M. Henry*[⋆] *Eleanor Hansen*[⋆] *Elizabeth Tandilashvili*[⋆]
*Lauren S. Wakschlag*[‡] *Elizabeth S. Norton*[¶] *Daniel S. Pine*[§] *Melissa A. Brotman*[⋆]
*Francisco Pereira*[†]

[⋆]Neuroscience and Novel Therapeutics Unit, National Institute of Mental Health
[‡]Department of Medical Social Sciences, Northwestern University
[¶]Department of Communication Sciences and Disorders, Northwestern University
[§]Section on Development and Affective Neuroscience Unit, National Institute of Mental Health
[†]Machine Learning Core, National Institute of Mental Health

## ABSTRACT

Infant crying can serve as a crucial indicator of various physiological and emotional states. This paper introduces a comprehensive approach detecting infant cries within audio data. We integrate Meta's Wav2Vec with traditional audio features, such as Mel-frequency cepstral coefficients (MFCCs), chroma, and spectral contrast, employing Gradient Boosting Machines (GBM) for cry classification. We validate our approach on a real-world dataset, demonstrating significant performance improvements over existing methods.

## 1. INTRODUCTION

Infant crying is a primary, evolutionarily conserved method of communication that signals needs and discomforts. Persistent crying can indicate distress in the infant and also cause stress in the caregiver. These negative emotions can adversely affect the caregiver's mental health and well-being, which in turn influences infant development and caregiver-infant dynamics. Recognizing these effects is crucial for conducting emotional research on both parents and children and for developing effective intervention strategies (e.g., see [1, 2]).

Traditionally, datasets are annotated by manually identifying crying events in audio data collected through continuous monitoring. The sheer volume of data makes manual annotation by researchers extremely time-consuming. Relying on parents to log crying times can introduce errors, as objective digital recordings often reveal that actual crying times are lower than parent-reported estimates, highlighting the limitations of subjective annotation (e.g., see [3]).

In many cases, naturalistic cry datasets must be annotated to be useful for scientific research. The large volume of data obtained through naturalistic collection often makes manual annotation infeasible. Automated annotation using machine learning offers a viable solution. By employing advanced algorithms, we can automate the detection and classification of crying events, significantly reducing human effort involved and enhancing the scalability of data analysis processes. This automation is crucial not only for operational efficiency but also for ensuring the accuracy and consistency of annotations, which are vital for subsequent analyses. Accurate annotations enable researchers to reliably quantify crying episodes and correlate them with clinical outcomes, providing deeper insights into infant health and caregiver-infant dynamics.

Our contribution is a novel approach that combines traditional and deep learning techniques to detect infant cries from audio recordings obtained from wearable devices. This approach significantly improves the accuracy and area aunder the curve (AUC) metrics of existing methods, thereby setting a new standard for automated crying detection.

## 2. RELATED WORK

Recent advancements in infant cry signal analysis have employed a combination of traditional machine learning classifiers and modern neural network techniques. Ji et al. [4] provide a comprehensive review of these advancements, covering topics from data acquisition to cross-domain signal processing. They emphasize the effectiveness of classifiers, including SVM [5] and CNN [6], and highlight the importance of feature extraction methods and the integration of neural network technologies, to enhance classification performance. Despite the rapid advancements in deep learning for acoustic signal processing, traditional methods like MFCC [7] still play a crucial role due to their interpretability, efficiency, and robustness, particularly in contexts with limited data (see also Bianco et al. [8] and Domingos et al. [9]).

For the specific problem of infant cry detection, Yao et al. [10] combined traditional acoustic features with deep spectrum features, derived from 2D mel-spectrograms using an AlexNet [11] model to classify crying. These features were then used as inputs to a linear SVM classifier. They achieved state-of-the-art performance on a dataset [12] released with their paper, which we will use for training and evaluation of our own approach.

The advent of Large Language Models (LLMs) like OpenAI's ChatGPT [13] and Meta's LLaMA [14] has transformed natural language processing and extended to domains like audio signal processing. Meta's wav2vec [15], initially developed for speech recognition, has been adapted for non-speech tasks such as sound event detection, acoustic scene classification, music tagging, and human action classification, achieving state-of-the-art results with reduced reliance on labeled data [16]. Wav2vec embeddings also address challenges like variable-duration utterances in audio data, enabling their use in more complex methods [17].

Our approach is inspired by Yao et al. [10] and uses the same traditional acoustic features. We replace their custom deep spectrum features with a general-purpose deep spectrum audio representation, wav2vec. Finally, we combine traditional and deep spectrum audio features as input to a GBM classifier. We chose a GBM due to its ability to model nonlinear interactions between features, and deliver high classification performance in speech tasks, as demonstrated by Dash et al. [18].

## 3. SYSTEM DESIGN

### 3.1. Feature space

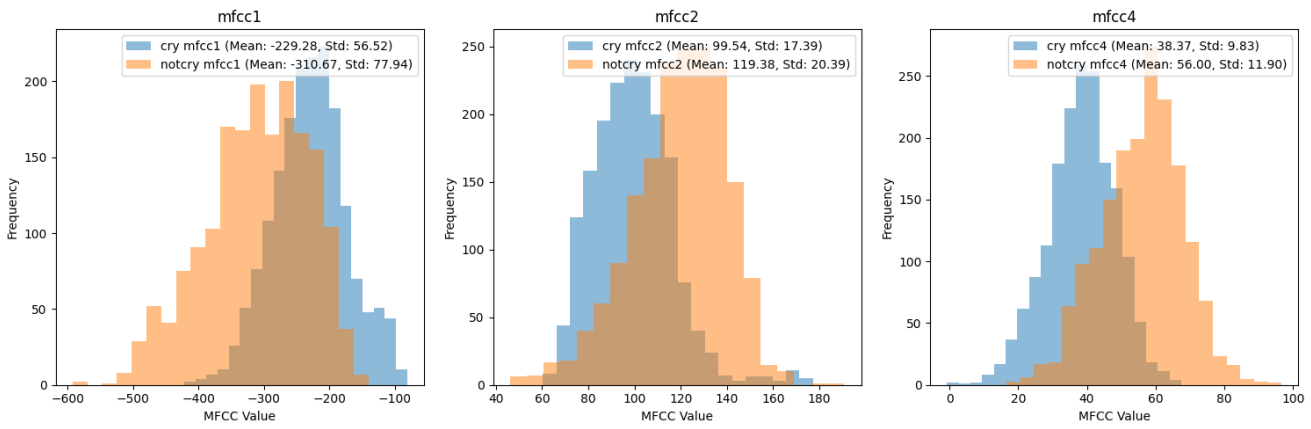

**Fig. 1**. Comparison of MFCC distributions for 'cry' and 'notcry' categories reveals that MFCC1, MFCC2, MFCC4, and MFCC5 exhibit differences in distributions, emphasizing their utility in distinguishing between crying and non-crying sounds.

The first component of our feature space consists of MFCC features, which are widely used in speech and audio processing due to their ability to capture the power spectrum of an audio signal in a compact form. Clear acoustic differences between 'cry' and 'not cry' training set samples are illustrated in Figure 1 for MFCC1, MFCC2, and MFCC4. MFCC1 captures the higher energy typical of crying sounds, while the means of MFCC2 and MFCC4 reflect differences in spectral shape. Therefore, MFCC features are likely to be informative for a cry classifier. The second component consists of Wav2Vec 2.0 features, which provide deep contextual information that complements the short-term power spectrum captured by MFCC features. Finally, the third component comprises spectral contrast and chroma features. Spectral contrast distinguishes signals by measuring amplitude differences between peaks and valleys [19], while chroma features represent the intensity of each of the twelve different pitch classes [20], adding a harmonic dimension that is vital for the comprehensive analysis of audio signals.

### 3.2. System Design

Figure 2 presents a flowchart contrasting our approach for crying classification with three different baselines, all of which operate on raw 5-second audio files. The first baseline is the approach from Yao et al. [10]. It generates a mel-spectrogram, which is then processed by an AlexNet model as a visual input, yielding a vector representation. In tandem, it extracts short-term traditional audio features. The two feature sets are concatenated and passed to an SVM classifier with an RBF kernel. The predicted label is a vote combining 5 per-second predictions of the classifier.

The cry sounds of human infants have been thought to contain only audible frequencies, with an average fundamental frequency of 300–600 Hz, until now [21]. Although this range is well-established, our method aims to avoid excluding potentially relevant higher frequencies while still filtering out high-frequency noise. Therefore, we applied a bandpass filter that retains frequencies between 300 Hz and 3000 Hz. These frequencies are particularly relevant for human vocal sounds and help filter out irrelevant noise, ensuring that subsequent analyses focus on the most informative parts of the signal.

The remainder of Figure 2 details our proposed approach, along with two baselines obtained by replacing the GBM classifier with a linear- or RBF-kernel SVM; every other step is shared among them. The initial stage involves preprocessing the audio data to enhance signal quality and isolate important characteristics, including checking for silence using a threshold-based method. After preprocessing, the system extracts a set of features combining Wav2Vec 2.0, traditional MFCCs, spectral contrast, and chroma features. The processed data then undergoes a final preparation phase where the selected features are standardized using a StandardScaler to ensure a mean of zero and a standard deviation of one, thereby preventing bias in model training due to differing scales of features. We use a StandardScaler within a pipeline to scale features during training, saving the learned scaling factors. During prediction, the pipeline reloads these trained scaling factors to ensure consistent scaling before feature selection and classification. The standardized data is subsequently used to train the GBM model.

GBM was chosen for its high performance across many domains, and robustness in handling complex datasets. It builds multiple decision trees sequentially, with each tree correcting the errors of its predecessor, leading to a strong predictive model. Once trained, the GBM model is used to predict and classify new instances. This involves applying the same preprocessing and feature extraction steps to new audio samples, followed by feature selection and scaling using the previously determined parameters. To evaluate the efficacy of the GBM, we also provide baselines that use the same preprocessing and feature extraction as our proposed framework, but replace the GBM with SVM-rbf and SVM-linear classifiers.

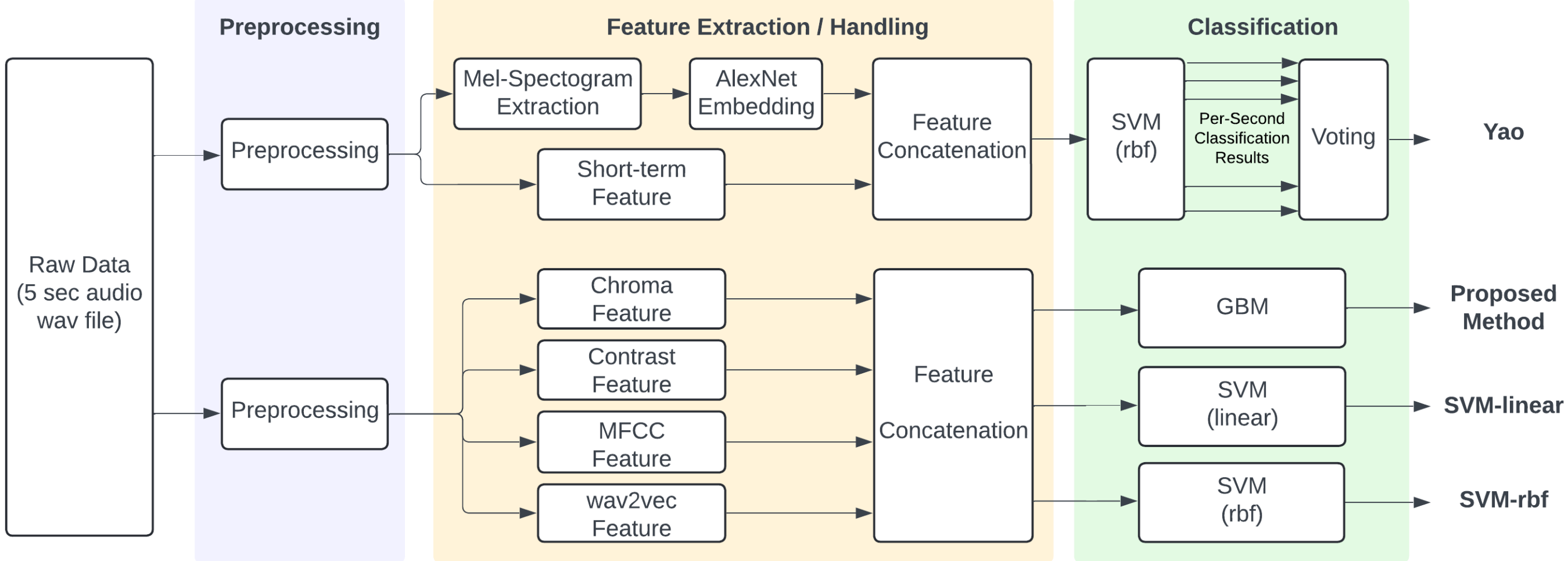

**Fig. 2**. Flowchart of Audio Classification Frameworks: Baseline and Proposed Approaches

## 4. EXPERIMENTS

### 4.1. Dataset Specification

For evaluation, we used the HomeBank English deBarbaro Cry Corpus [12], which includes data from the RW-Filt dataset. This dataset, featuring annotated recordings of infant vocalizations with a focus on early developmental sounds, is well-suited for our analysis. It consists of 5-second audio segments labeled as cry/not-cry in roughly equal proportions, collected from 21 infants wearing recording vests. The data was split into training and testing subsets with distinct participants: 15 for training (25,717 segments) and 6 for testing (14,812 segments). This setup ensures a robust evaluation of the model's ability to classify short audio samples and generalize to new participants.

### 4.2. Model Comparison

The primary comparison is between our proposed GBM approach and the current state-of-the-art method, proposed by Yao et al. [10]. For simplicity, we will refer to this method as 'Yao.' The other two baseline approaches are identical to GBM but use linear and RBF-kernel SVM classification models instead. We re-implemented the Yao approach to allow for re-training on new data, by re-using their code for preprocessing and feature extraction, and developing PyTorch implementations of their original TensorFlow code with the same parameter choices. The classification model in this approach produces predictions at a 1-second level granularity, which are then combined through a voting mechanism into a label for the full 5-second audio segment. For a fair benchmark comparison, we also did not tune any parameters in our approaches and simply used the default values. This ensures that all approaches have room for improvement when applied to new datasets.

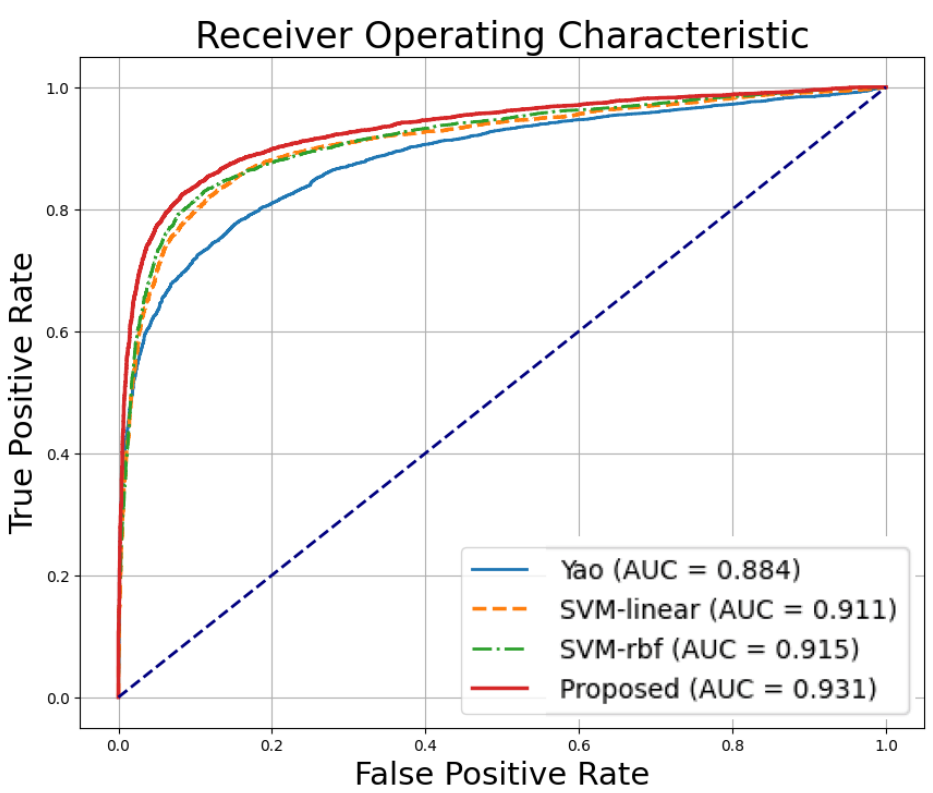

**Fig. 3**. Receiver Operating Characteristic (ROC) curves comparing the performance of different classification methods.

### 4.3. Experimental Results

Figure 3 presents the receiver operating characteristic (ROC) curves for each approach on the test set, along with their respective areas under the curve (AUC). The ROC curves visually represent the trade-offs between true positive rates and false positive rates. Depending on the downstream application of the model predictions, the optimal trade-off between these rates may differ. Given that, we believe the AUC is the most appropriate performance measure. Our method achieves the highest AUC of 0.931, followed by SVM-RBF (0.915) and SVM-linear (0.915) methods, and then Yao (0.884).

| | Proposed | Linear | RBF | Yao | # examples |
|---|---|---|---|---|---|
| P26 | **0.9559** | 0.9111 | 0.9118 | 0.8931 | 2666 |
| P15 | **0.9595** | 0.9398 | 0.9359 | 0.9406 | 4052 |
| P38 | 0.9499 | **0.9594** | 0.9586 | 0.9186 | 3544 |
| P36 | **0.9582** | 0.9405 | 0.9394 | 0.9167 | 2746 |
| P20 | **0.9532** | 0.9522 | 0.9498 | 0.8948 | 2738 |
| P22 | 0.8682 | 0.8516 | **0.8708** | 0.8414 | 2666 |
| Overall | **0.9408** | 0.9258 | 0.9277 | 0.9009 | 18412 |

**Table 1**. AUC of ROC Curve

| | Proposed | Linear | RBF | Yao |
|---|---|---|---|---|
| Proposed | - | 0.9017 | 0.7602 | 1.0 |
| Linear | 0.0983 | x | - | - |
| RBF | 0.2398 | - | x | - |
| Yao | 0.0 | - | - | x |

**Table 2**. Pairwise Probabilities of Model Superiority using Bayesian Comparison. Statistically indistinguishable performance is noted with $-$.

Table 1 shows the AUC of ROC curves calculated separately for each test participant, highlighting the consistent advantage of our approach over the Yao baseline. To formalize this observation, we performed a Bayesian comparison of model performance across test participants using the Baycomp software [22], which provides a robust framework for comparing classifiers through probabilistic inference. Table 2 presents pairwise probabilities of model superiority, where rows and columns represent the models being compared. For instance, the cell for GBM vs. Linear (0.9017) indicates a 90.17% probability that GBM is superior, while GBM vs. Yao shows a 100% probability of GBM outperforming the Yao model. These results strongly suggest that GBM outperforms the other models.

**Table 3**. Classification Report for Different Methods

| Method | Metric | Class 0 | Class 1 | Macro Avg | Weighted Avg | Accuracy |
|---|---|---|---|---|---|---|
| Proposed | Precision | 0.845 | 0.897 | | | |
| | Recall | 0.904 | 0.834 | 0.871 | 0.871 | 0.869 |
| | F1-Score | 0.873 | 0.864 | | | |
| | Support | 5408 | 5408 | 10816 | 10816 | |
| SVM-linear | Precision | 0.818 | 0.887 | | | |
| | Recall | 0.898 | 0.800 | 0.852 | 0.852 | 0.849 |
| | F1-Score | 0.856 | 0.841 | | | |
| | Support | 5408 | 5408 | 10816 | 10816 | |
| SVM-rbf | Precision | 0.825 | 0.897 | | | |
| | Recall | 0.907 | 0.808 | 0.861 | 0.861 | 0.857 |
| | F1-Score | 0.864 | 0.850 | | | |
| | Support | 5408 | 5408 | 10816 | 10816 | |
| Yao | Precision | 0.837 | 0.767 | | | |
| | Recall | 0.740 | 0.855 | 0.802 | 0.802 | 0.798 |
| | F1-Score | 0.785 | 0.809 | | | |
| | Support | 5408 | 5408 | 10816 | 10816 | |

Finally, we calculated additional measures of performance to make comparison with other detection studies easier, and summarized them in Table 3. Note that these measures are driven by accuracy, which determines a particular precision/recall trade-off, in contrast with the AUC. Overall, if optimizing for accuracy, our approaches outperform Yao.

### 4.4. Ablation Analysis

**Table 4**. Feature Contribution Results

| Feature Set | Accuracy | F1-Score |
|---|---|---|
| Wav2Vec | 0.852 | 0.848 |
| MFCCs | 0.848 | 0.849 |
| Chroma | 0.688 | 0.686 |
| Spectral Contrast | 0.768 | 0.762 |
| Combined | **0.873** | **0.874** |

To identify the most significant contributors to model performance, we systematically evaluated individual feature sets, as summarized in Table 4. Wav2Vec features achieved an accuracy of 0.852 and an F1-score of 0.848, while MFCCs reached 0.848 and 0.849, respectively. Chroma features showed lower performance, with 0.688 accuracy and 0.686 F1, and spectral contrast yielded 0.768 accuracy and 0.762 F1. The combined feature set achieved the highest performance (0.873 accuracy, 0.874 F1), demonstrating the complementary strengths of Wav2Vec and MFCCs. Despite some redundancy, integrating all feature types provided the best results.

## 5. CONCLUSION

Our proposed method employs a GBM using features extracted from Wav2Vec, MFCCs, chroma, and spectral contrast, achieving substantial AUC improvements over the state-of-the-art method by Yao et al. [10]. Comparison with linear- and RBF-kernel baselines attributes this performance boost primarily to the feature space and secondarily to the GBM. Notably, all feature types can be generated automatically without parameter tuning or additional model training, making complex audio classification problems more accessible. These advancements could pave the way for improved diagnostic tools in pediatric healthcare, enabling better-informed caregiving decisions and fostering enhanced early childhood health and well-being.

## 6. ACKNOWLEDGEMENTS

This research was supported by the Intramural Research Program of the National Institute of Mental Health (NIMH), under projects ZIC-MH002968 (PI: Dr. Francisco Pereira) and ZIA-MH002969 (PI: Dr. Melissa A. Brotman).